\documentclass[final]{agujournal2019}
\usepackage{url} 
\usepackage{lineno}
\usepackage[inline, ignoremode]{trackchanges} 
\usepackage{soul}
\draftfalse

\journalname{JGR: Space Physics}

\begin{document}

%
%




\title{Jovian sodium nebula and Io plasma torus S$^+$ and brightnesses
  2017 -- 2023: insights into volcanic vs.\ sublimation supply}



%
%




\authors{Jeffrey P.\ Morgenthaler\affil{1},
  Carl A.\ Schmidt\affil{2},
  Marissa F.\ Vogt\affil{1,2},
  Nicholas M.\ Schneider\affil{3},
  Max Marconi\affil{1}}

\affiliation{1}{Planetary Science Institute 
  1700 East Fort Lowell, Suite 106
  Tucson, AZ 85719-2395, USA}
\affiliation{2}{Center for Space Physics
  Boston University
  Boston, MA 02155, USA}
\affiliation{3}{University Of Colorado, Boulder
  Boulder, CO 80309, USA}




\correspondingauthor{Jeffrey P.\ Morgenthaler}{jpmorgen@psi.edu}



\begin{keypoints}

\item A large set of Jovian sodium nebula and Io
  plasma torus S$^+$ images provides context for Io and Jovian
  magnetospheric studies
\item Enhancements in Na and S$^+$ emission last 1 -- 3 months, ruling out
    insolation-driven sublimation as their driver
\item Volcanic plumes likely play a key role in atmospheric escape

  
%
%
%
\end{keypoints}

%
%

%
%


\begin{abstract}

We present first results derived from the largest collection of
contemporaneously recorded Jovian sodium nebula and Io plasma torus
(IPT) in [S\;II]\,6731\,\AA\ images assembled to date.  The data were
recorded by the Planetary Science Institute's Io Input/Output
observatory (IoIO) and provide important context to Io geologic and
atmospheric studies as well as the \textsl{Juno} mission and
supporting observations.  Enhancements in the observed emission are
common, typically lasting 1 -- 3 months, such that the average flux of
material from Io is determined by the enhancements, not any quiescent
state.  The enhancements are not seen at
  periodicities associated with modulation in solar insolation of Io's
  surface, thus physical process(es) other than insolation-driven
  sublimation must ultimately drive the bulk of Io's atmospheric
  escape.  We suggest that geologic activity, likely involving
  volcanic plumes, drives escape.

\end{abstract}


\section*{Plain Language Summary}

The Planetary Science Institute's Io Input/Output observatory (IoIO)
is composed almost entirely of off-the-shelf parts popular with
amateur astronomers.  IoIO uses special filters to isolate emission
from two gasses found around Jupiter: neutral sodium and ionized
sulfur.  The sodium is thrown out from Io in a vast cloud called the
Jovian sodium nebula.  The ionized sulfur collects into the Io plasma
torus (IPT), a ring-shaped structure centered around Jupiter that
wobbles around Io's orbital path.  These gasses ultimately come from
Jupiter's highly volcanic moon, Io.  We see the Na nebula and IPT
brighten frequently.  This demonstrates that the majority of the
material leaving Io comes from whatever drives the frequent
brightening events, with volcanic plumes
likely playing a key role.  Our results challenge a widely held
belief in the scientific community, that the majority of the material
in the Na nebula and IPT comes from Io's tenuous global atmosphere,
which is fed by the sublimation of surface frosts and is relatively
stable in time.  Our dataset also provides important context for
NASA's \textsl{Juno} mission and supporting observations that focus on
Io volcanism, the material's likely source, and Io's magnetosphere,
the material's ultimate destination.

\section{Introduction}
\label{introduction}

One of the first hints that Io was somehow releasing material into the
Jovian magnetospheric environment in large amounts compared to the
other Galilean satellites came from spectroscopic observations of
sodium D1 (5896\,\AA) and D2 (5890\,\AA) emissions that were time
variable, broad, and in a ratio suggesting optically thick gas
\cite{brown74}.  \citeA{kupo76} conducted spectroscopic studies of
S$^+$ in the [S\;II] 6717\,\AA\ and 6731\,\AA\ doublet in the orbital
plane of Io and detected extended emission corotational with Jupiter.
As \textsl{Voyager I} approached Jupiter, Extreme ultraviolet (EUV)
emission from S\,III, S\,IV, and O\,III resolved into a torus-like
structure encircling Jupiter, dubbed the Io plasma torus
\cite<IPT;>{broadfoot79}.  The potential source of the material
  escaping Io was first hinted at when Io itself was
also seen to be intermittently bright at a particular orbital phase
angle in the 3 -- 5\,$\mu$m region of the infrared spectrum, with
volcanic activity being one of several possible
explanations offered \cite{witteborn79}.  Volcanic activity on Io was
subsequently unambiguously confirmed by \textsl{Voyager}~1 images of
plumes and volcanic surface features \cite{morabito79, smith79}.  This
volcanism gives rise, either directly or indirectly, to an SO$_2$
dominated atmosphere \cite{pearl79, kumar79, depater21} with minor
constituents, including NaCl \cite{lellouch03, mcgrath04, moullet10,
  redwing22}.

Io's atmosphere undergoes several reactions with material in the IPT,
resulting in detectable effects.  For majority species S and O,
change exchange, sputtering, and electron impact
  ionization are important processes for removing atmospheric
material \cite<e.g.,>{mcgrath87, smyth88a, smyth88, thomas04,
  schneider07, dols08, dols12, smith22}, resulting in a roughly
torus-shaped neutral cloud confined to Io's orbital plane and mapped
in the EUV at O\;I 1304\,\AA\ \cite{koga18_space}.  IPT electron
impact ionization of this neutral cloud is the primary process by
which the IPT receives new material, with direct ionization in Io's
atmosphere providing only a minor component \cite{dols08, dols12}.
The canonical value of $\sim$1\,ton\,s$^{-1}$ of material flowing into
the IPT from Io has been estimated using the IPT's total EUV power
output and a simple geometric model of the EUV emission region
\cite<e.g.,>{broadfoot79, schneider07}.

The path that sodium-bearing material takes as it escapes Io's
atmosphere is different, thanks to the low ionization potential of any
sodium-containing molecule, NaX.  For these molecules, impact
ionization and charge exchange processes are very efficient
\cite<e.g.,>{schneider07}.  Pickup NaX ions generated in Io's
exosphere that promptly neutralize and dissociate create a directional
feature, the ``jet,'' that points radially outward from Io and flaps
up and down in synchrony with the IPT \cite[see also animations
  accompanying Figure~\ref{fig:Na_SII_images}]{pilcher84, wilson99,
  burger99}.  A more extended structure, dubbed the ``stream,'' has a
similar radial morphology and behavior relative to IPT modulation, but
extends for several hours in Jovian local time downstream of Io's
position.  This comes from NaX$^+$ ionized in Io's exosphere by IPT
plasma and swept downstream in the plasma flow before they neutralize
\cite{schneider91Corona, wilson94}.  IPT NaX$^+$ ions that dissociate
produce neutral fragments that are ejected from the IPT at an average
of $\sim$70\,km\,s$^{-1}$, which is the Jovian corotational velocity
at the IPT.  This velocity is above Jupiter's escape velocity.  Thus,
neutral Na is detected at distances $>$500 Jovian radii
(R$_\mathrm{j}$) from Jupiter \cite{mendillo90}.  All these neutral
sodium emission features are known collectively as the Jovian sodium
nebula and are well-described by Monte Carlo modeling techniques
\cite{wilson02}.  The Jovian sodium nebula has been the subject of
several long-term studies using ground-based coronagraphic techniques
\cite[and this work]{mendillo04, yoneda09, yoneda15, roth20,
  morgenthaler19}.

Except for the study presented here, long-term observations of the IPT
from ground-based observatories have been limited.  The longest
continuous study to date covered a full Jovian opposition using a
spectroscopic technique \cite{brown97}.  Ground-based IPT imaging
campaigns using coronagraphic techniques have typically lasted a few
weeks per opposition, though some have extended over several
oppositions \cite{schneider95, woodward00_no_SII, nozawa04_no_SII,
  kagitani20}.  These ground-based observations concentrate on the
bright [S\;II] emissions of the 6717\,\AA, 6731\,\AA\ doublet, which
are excited by IPT thermal electrons.  A significant amount of
structure is seen in high-resolution IPT images: a dense ``ribbon''
near Io's orbit is separated by a gap from the more disk-like ``cold
torus,'' closer to Jupiter \cite<e.g.>{schneider95}.  There is
evidence that diffusion proceeds inward from the ribbon to the cold
torus \cite{herbert08}.  Long-term spectro-imaging observations of EUV
emission of the IPT have been conducted from \textsl{Voyager},
\textsl{Cassini}, and \textsl{Hisaki} \cite{broadfoot79, steffl04a,
  yoshioka14}.  This emission, known as the ``warm torus,'' is more
extended radially and vertically than the ribbon.  The EUV emissions
are excited by suprathermal electrons and have been used to study
radial transport in the IPT, providing evidence that the total
residence time of material in the IPT is 20 -- 80 days
\cite{bagenal11, hess11, copper16, tsuchiya18}.  Once material has
left the IPT, it rapidly spirals outward in a few days
\cite{bagenal11}.  Thus, long-term monitoring of the IPT, such as that
presented here, provides critical context to any study of Jupiter's
broader magnetosphere, such as that conducted by NASA's
\textsl{Juno} mission \cite{bolton17} and supporting observations
\cite{orton_planets2020, orton22}.

In this work, we present the first results of a combined Jovian sodium
nebula and IPT monitoring campaign, conducted since March 2017 by
  the Planetary Science Institute's Io Input/Output observatory
  (IoIO).  The coronagraphic observations are described in
\S\ref{observations} and \S\ref{reduction} provides the methodology
used to reduce the data.  Section \ref{results} presents the primary
results of our study, which is a time history of the surface
brightnesses of the Na nebula and IPT (Figure
\ref{fig:time_sequence}).  This time history shows 1 -- 2 brightness
enhancements per 7-month observing season, each lasting 1 -- 3 months,
such that emissions are seldom found in a quiescent state.  In
\S\ref{compare} we compare our results to previous studies, noting
that although none of the previous workers reported such frequent
activity in the Na nebula and IPT, all are consistent with it.
In \S\ref{discussion}, we use the IoIO data to rule
  out solar insolation-driven sublimation of Io's surface frosts as
  the primary driver of material from Io's atmosphere, showing instead
  that geologic processes must be involved.  We then review the
  existing evidence that connects enhancements in material escape from
  Io's atmosphere with volcanic plume activity and discuss
  implications for the transport of material.     A summary and concluding remarks
are provided in \S\ref{conclusion}.  In \S\ref{future}, we
suggest additional uses for the IoIO dataset, including providing
  support for  current and planned missions to Jupiter.

\section{Observations}
\label{observations}

All observations presented here were conducted with the Planetary
Science Institute's Io Input/Output observatory (IoIO).  IoIO consists
of a 35\,cm Celestron telescope feeding a custom-built coronagraph,
described by \citeA{morgenthaler19}.  Since the publication of that
work, both the observatory hardware and control software have been
upgraded, enabling fully robotic acquisition of Jovian sodium nebula
and IPT [S\;II] on- and off-band images, regular photometric
observations of \citeA{burnashev85} spectrophotometric standard stars
in all filters, and observations of telluric sodium foreground
emission.  Bias, dark and sky flat images are also periodically
recorded.  Since 2017-03-09, IoIO has contemporaneous recorded
Na\,5890\,\AA\ nebula and IPT [S\;II]\,6731\,\AA\ observations on over
550 nights, with over 2300 Na images and over 8300 [S\;II] images
collected.  The observatory has been operated on another $\sim$500
nights in support of other Planetary Science Institute projects
\cite<e.g.,>{adams23DPS} and
pilot studies, increasing the number of spectrophotometric
calibrations and time coverage of telluric Na emission, which provides
a time-variable foreground emission \cite<e.g.,>{plane18}.

\section{Data Reduction}
\label{reduction}

\subsection{General Considerations}
\label{reduction:general}

All IoIO data are reduced pipeline-style using the software enumerated
in the Open Research section.  The \citeA{burnashev85}
spectrophotometric observations show each filter in our filter library
provides stable zero points and extinction coefficients over the
length of our study, modulo random nightly variations due to variation
in atmospheric transparency, which we ignore in our current analyses,
using instead the biweight location \cite{tukey77}, of all the
measurements of each filter.  The biweight location, $\zeta_{biloc}$,
is defined as:

\begin{equation}
  \zeta_{biloc} = M + \frac{\sum_{|u_i|<1} \ (x_i - M) (1 - u_i^2)^2}
       {\sum_{|u_i|<1} \ (1 - u_i^2)^2}
\label{eq:biweight}
\end{equation}

\noindent
where $x$ is the data, $M$ is the median of the data.  The quantity
$u_i$ is:

\begin{equation}
        u_{i} = \frac{(x_i - M)}{c * MAD}
\label{eq:biweight_u}
\end{equation}

\noindent
where $c$ is a tuning constant, set to 9 in our case, and $MAD$ is the
median absolute deviation:

\begin{equation}
        MAD = \mathrm{median}(|(x_i - \overline{x}|)
\label{eq:MAD}
\end{equation}

\noindent
The biweight location is a more robust statistical measure of the
central location of a distribution than the median, particularly for
data not distributed as a Gaussian \cite{beers90}.  Surface
brightnesses are expressed in rayleighs (R), where:

\begin{equation}
	1~R =
        {10^6\over4\pi}~\mathrm{photons~cm^{-2}~sec^{-1}~sr^{-1}}
\label{eq:rayliegh}
\end{equation}

\begin{figure}
  \noindent\includegraphics[width=0.9\textwidth]{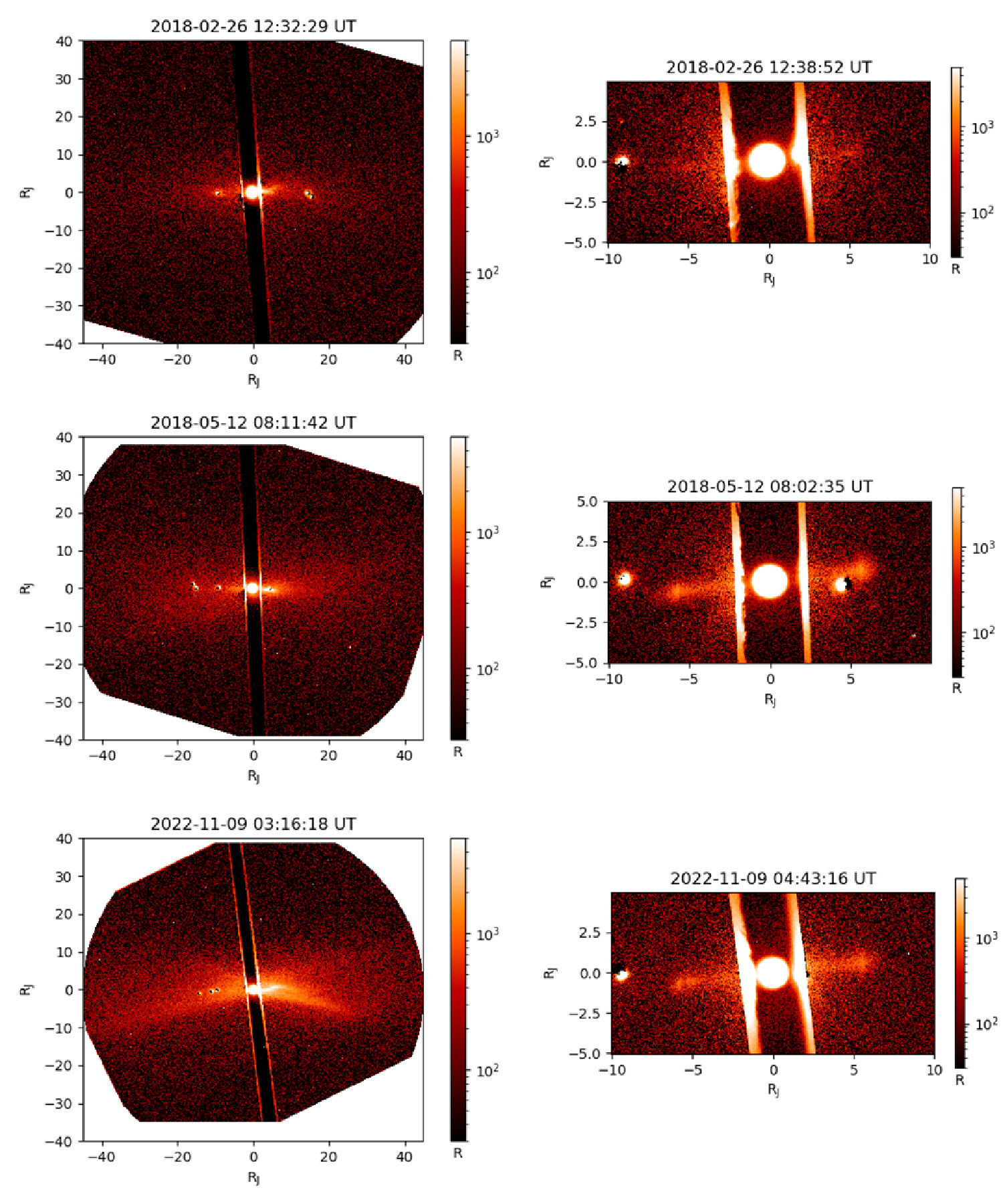}
\caption{Sample IoIO Na nebula (left column) and IPT
  [S\;II]\,6731\,\AA\ (right column) images.  The top row shows images
  recorded just before the large 2018 enhancements in Na nebula and
  IPT emission (see Figure \ref{fig:time_sequence}), the middle row
  images recorded near the 2018 Na nebula and IPT peaks in emission
  and the bottom row images recorded near the 2022 peaks.  Animations
  are provided in the on-line journal.}
\label{fig:Na_SII_images}
\end{figure}

Astrometric solutions of our images, together with high-quality JPL
HORIZONS ephemerides \cite{giorgini96} enable high-precision alignment
of on- and off-band images before subtraction of the off-band images.
Subtraction of the off-band images effectively removes Jupiter's
scattered continuum light from the on-band images.  When astrometric
solutions using field stars fail, the position of Jupiter on the
coronagraph neutral density (ND) filter is used to establish the
astrometric center of the image, with the clock angle determined by
the previous successfully solved image.  As expected from our stellar
calibrations, we found the ratio between our on- and off-band sky
flats gave stable results over the lifetime of the project.  Thus, we
used the biweight location of all ratios to scale the off-band images
before subtraction from the on-band.  Sample reduced images are shown
in Figure \ref{fig:Na_SII_images}.

%
%

%

We note that our calibration procedure is a significant improvement
over the technique used by \citeA{morgenthaler19}, which relied on the
image of Jupiter through the ND filter for surface brightness
calibration.  As discovered after the installation of a larger and
filter wheel in 2019, Jupiter's detected brightness is subject to an
unexpected Fabry P\'erot effect between the narrow-band and ND
filters, with each narrow-band filter providing a different magnitude
of effect.  Our current procedure avoids this issue by using the
stellar and flat-field calibrations described above.

In order to establish a time-sequence of the Na nebula and IPT
brightnesses, we first rotate the images reduced by the
  procedure above into the plane of the IPT centrifugal equator using
the relation:

\begin{equation}
\alpha = -A \times\ cos(\lambda_{\mathrm{III}} - P)
\label{eq:IPT_NPole_ang}
\end{equation}

\noindent
where $\alpha$ is the angle between the Jovian rotational axis and the
perpendicular to the IPT centrifugal equator, $\lambda_{\mathrm{III}}$
is the sub-observer System III longitude, $A$ is the amplitude of the
oscillation of the centrifugal equator and $P$ is the
$\lambda_{\mathrm{III}}$ longitude of the intersection of the magnetic
and equatorial planes.  For this work, we used $A$ = 6.8$^{\circ}$
\cite{moirano21IPT} and $P$ = 290.8$^{\circ}$ \cite{connerney98}.
Values of $A$ = 6.3$^{\circ}$ \cite{phipps21_equator} and $P$ =
286.61$^{\circ}$ \cite{connerney18} could also be used, and would
result in trivial differences in our extracted surface brightnesses.

\subsection{Na Nebula}
\label{reduction:Na}

\begin{figure}
 \noindent\includegraphics{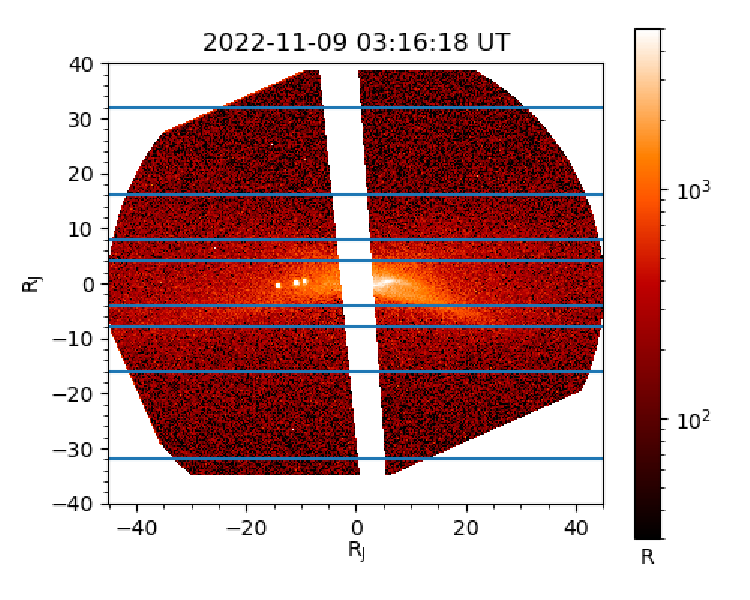}
\caption{Sample Na nebula image illustrating reduction steps described
  in \S\ref{reduction:Na}.  The blue lines indicate boundaries between
  apertures used to extract surface brightness values as a function of
  vertical distance from the IPT centrifugal plane.  The boundaries
  between apertures are defined by the following vertical distances
  from the IPT centrifugal plane: 4\;R$_\mathrm{j}$ --
  8\;R$_\mathrm{j}$, 8\;R$_\mathrm{j}$ -- 16\;R$_\mathrm{j}$, and
  16\;R$_\mathrm{j}$ -- 32\;R$_\mathrm{j}$, with the average distances
  from the plane of each pair of apertures used for subsequent
  identification (e.g., see legend of Figure \ref{fig:time_sequence},
  top).  Masked areas are shown in white.}

\label{fig:Na_apertures}
\end{figure}

As shown by previous work (\S\ref{introduction}) and the Na nebula
animations accompanying Figure \ref{fig:Na_SII_images} in the online
journal, the bulk of the bright jet and stream emission follow Io in
its 42\,hour orbit and flap up and down with each 9.925\,hr Jovian
rotation.  To minimize the effects of this high variability when
extracting surface brightnesses from individual Na nebula images, we
rotate each image by $\alpha$, as described above, and divide
the resulting image into horizontal apertures distributed vertically
from the IPT centrifugal plane, as shown in Figure
\ref{fig:Na_apertures}.  The ND filter and area beyond the edge of the
narrow-band filters are masked, as are pixels with values above the
non-linear point of the CCD, with a larger mask area applied around
Galilean satellites.  The average surface brightness in each aperture
is calculated by totaling the individual surface brightnesses of the
unmasked pixels and dividing by the total number of unmasked pixels.
The final surface brightness for a given distance from the IPT
centrifugal plane is the average of the surface brightnesses of the
pair of apertures located at that distance above and below the plane.

\subsubsection{Removal of Telluric Sodium Contamination}
\label{reduction:telluric_Na}

Telluric sodium emission provides a time-varying and, at times,
substantial field-filling component to our Na nebula images.  We
attempted to remove this emission using an empirical model constructed
from our multi-year dataset of telluric sodium emission observations.
The model accounted for airmass effects, solar scattering angle, and
seasonal effects.  However, after subtraction of the model, the time
sequence of Na nebula surface brightnesses was still quite noisy.
Thus, we instead subtract the average surface brightness of emission
$>$32\;R$_\mathrm{j}$ above and below the centrifugal plane from the
extracted surface brightnesses of each image.  As a result, the
variation induced by telluric emission was greatly diminished.  A
final step in the Na nebula reduction is to compute the biweight
location of all of the measurements at each distance on each night.
The results are plotted together with 21-point moving median filter in
Figure \ref{fig:time_sequence} (top).

A byproduct of our telluric sodium removal technique is to induce an
intensity-dependent error of the order $\sim$ 5\,R -- $\sim$ 25\,R in
our quoted Na nebula surface brightnesses.  This is because our
telluric removal procedure effectively assumes that the brightness of
the Na nebula is zero at the edge of the IoIO FOV.  However, as shown
by larger-field images \cite{mendillo04, yoneda09, yoneda15}, the
emission $>$30\;R$_\mathrm{j}$ above and below Jupiter's equatorial
plane varies from $<$ 5\,R to $\sim$ 25\,R, depending on whether or
not the nebula is enhanced.  This effect could be corrected using a
model of the Na nebula emission, however, doing so would not affect
the results of our current study.

\subsection{IPT}
\label{reduction:IPT}

\begin{figure}
 \noindent\includegraphics[width=0.9\textwidth]{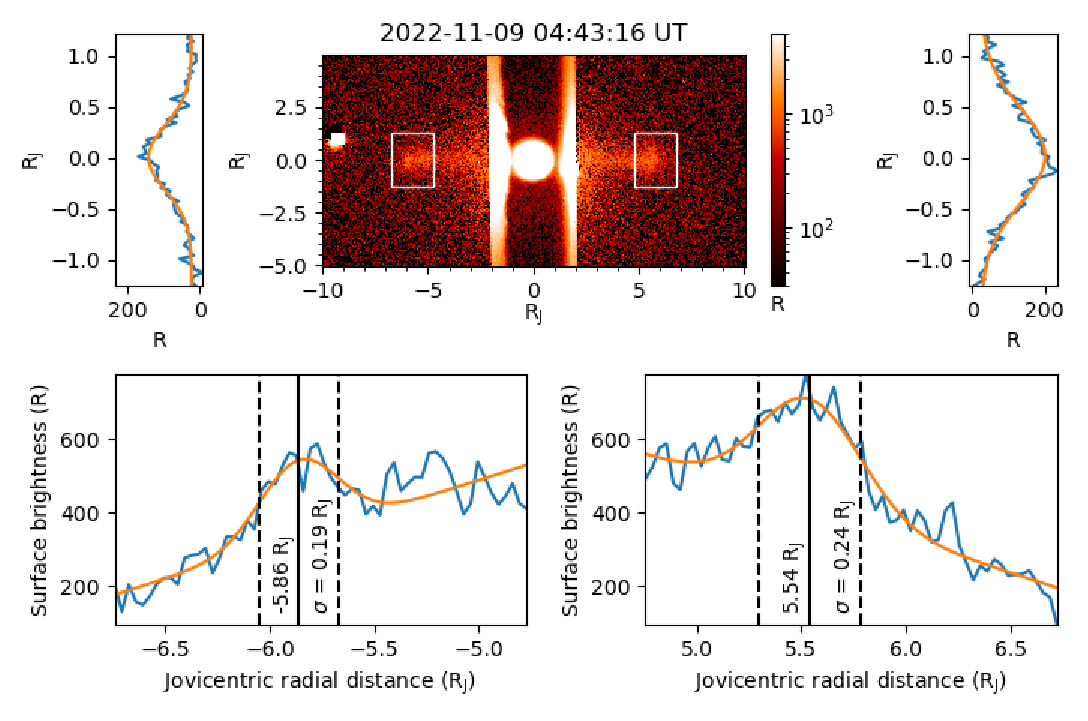}
\caption{Graphical depiction of [S\;II] ribbon surface brightness
  extraction process described in \S\ref{reduction:IPT}.  An IPT image
  rotated into the reference frame of the IPT centrifugal equator is
  shown in the top, middle panel.  Radial profiles of the ansas are
  shown in the bottom row, fit by Equation\,\ref{eq:ansa_rad}.  The
  1-$\sigma$ limits indicated on these plots define the edges of
  regions between which the vertical profiles are computed (top row,
  outer plots).  The average surface brightness of each ribbon is the
  integral of the Gaussian component of the fit of the corresponding
  vertical profile.}

\label{fig:SII_reduction}
\end{figure}

\begin{sidewaysfigure}
    \noindent\includegraphics[width=\textheight]{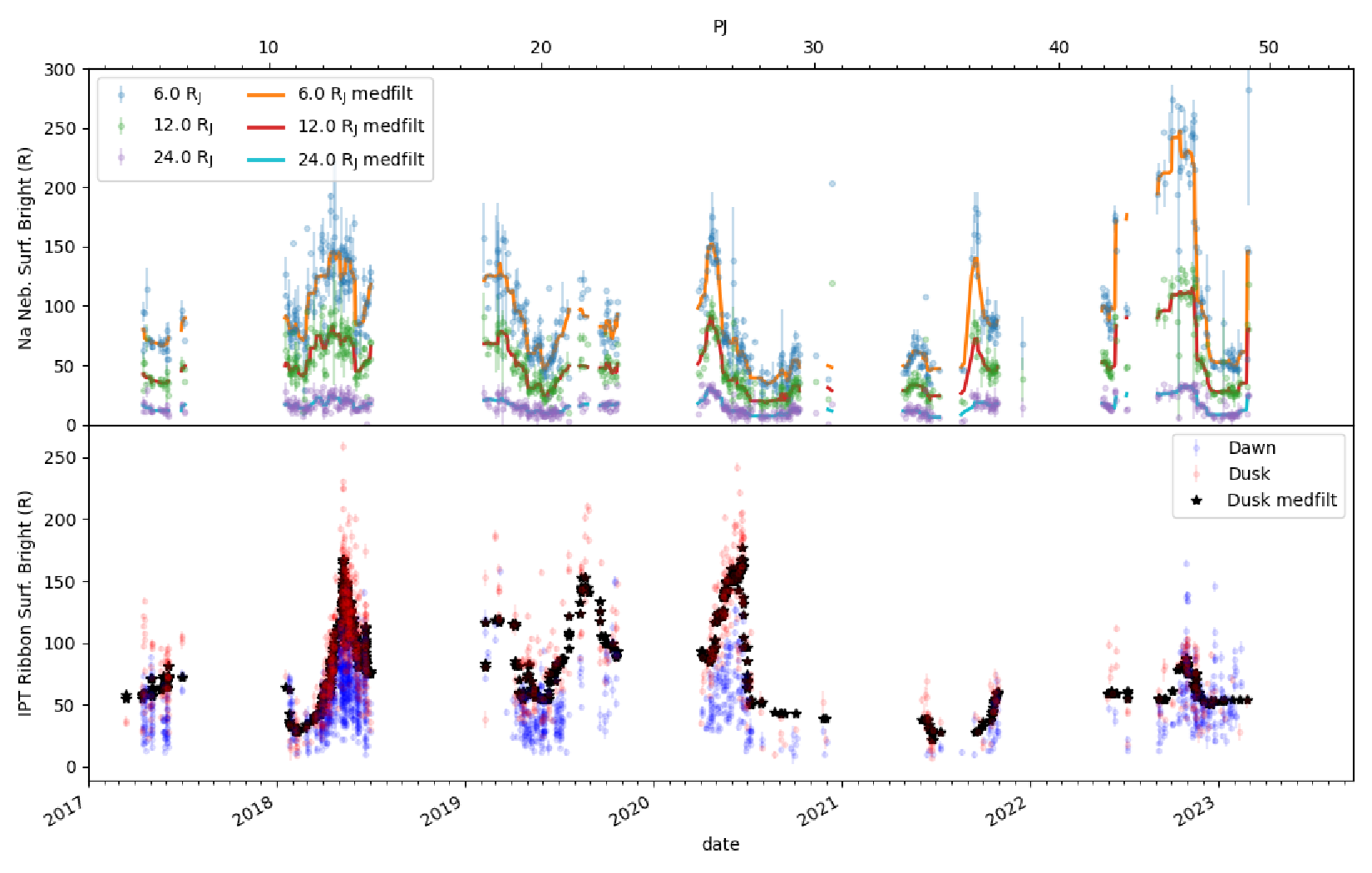}

\caption{Top: Time sequence of surface brightnesses in Jovian sodium
  nebula at three distances from the plasma torus equatorial plane.
  Bottom: Time sequence of IPT ribbon average surface brightnesses.  A
  running median filter, 21 points wide, is applied to each Na
  aperture and the dusk ribbon brightnesses.}

\label{fig:time_sequence}
\end{sidewaysfigure}

As shown by comparing Figure \ref{fig:Na_SII_images} (lower right) to
the IPT image in Figure \ref{fig:SII_reduction}, rotating by Equation
\ref{eq:IPT_NPole_ang} provides a natural coordinate system for
extracting brightness values of the ansas (edges, Latin: ``handles'')
of the IPT.  As described in \S\ref{introduction}, the
[S\;II]\,6731\,\AA\ ansas primarily capture the IPT ribbon emission.
We extract the average surface brightness from each ribbon feature
using the two-step process shown graphically in Figure
\ref{fig:SII_reduction}.  Specifically, starting from the rotated
image, we define ansa extraction regions that extend radially
4.75\;R$_\mathrm{j}$ to 6.75\;R$_\mathrm{j}$ from Jupiter and
$\pm$1.25\;R$_\mathrm{j}$ above and below the IPT centrifugal equator
(white boxes).  Radial profiles of the emission in the white boxes are
shown in the bottom row.  These profiles are generally well-fit by a
Gaussian plus sloping continuum of the form:

\begin{equation}
f(x) = A e^{-{(x-x_0)^2}\over{2\sigma^2}} + P_0 + P_1x
\label{eq:ansa_rad}
\end{equation}

\noindent
where $A$ is the peak surface brightness of the Gaussian component of
the radial profile, $x_0$ is the ribbon radial distance from Jupiter,
$\sigma$ is the width of the ribbon, and $P_0$ and $P_1$ are the
coefficients of the linear background.  The $\pm$1-$\sigma$ limits of
these Gaussians are used to define the radial limits of the region
used to extract vertical profiles, shown in the top row, outside
plots.  Equation \ref{eq:ansa_rad}, with $P_1$ = 0, is used to fit the
vertical profiles.  The Gaussian component of this function is
integrated to arrive at an average ribbon intensity of
$A\sigma\sqrt{2\pi}$.  This is converted to an average surface
brightness by dividing by $\sigma$.  Occasionally, the data are of
high enough quality and the torus configured such that cold torus is
resolved.  This is the case for the dawn (left) ansa and results in a
small peak inside of the ribbon.  We ignore the effect of this feature
on our fits, since the simple sloping continuum plus Gaussian
provides an adequate a foundation for determining the region over
which to extract vertical profiles.  As shown in the Figure, the
vertical profiles are well-described by Equation \ref{eq:ansa_rad},
with $P_1$ = 0.  The total area of the Gaussian components of each fit
is then used to establish the average surface brightness of each
ribbon.  If an extraction area contains saturated pixels from any
nearby Galilean satellite, it is excluded from the analysis.  Fits
that result in ribbon peak positions outside of the range
5.0\;R$_\mathrm{j}$ -- 6.5\;R$_\mathrm{j}$ or peak widths outside the
range 0.1\;R$_\mathrm{j}$ -- 0.5\;R$_\mathrm{j}$ are discarded.  In
this way, our extractions are able to adjust for varying observing
conditions and the intrinsic variability in the IPT ansa morphology
\cite<e.g.,>{schneider95} and reliably discard pathological cases.
The time history of the average ribbon surface brightnesses, together
with a 21-point running median filter of the dusk ribbon points are
plotted in Figure \ref{fig:time_sequence} (bottom).  On timescales of
weeks to months, all other parameters of the fits roughly scale with
ribbon surface brightness, except for the radial peak positions of the
ribbon.  This behavior is expected because all of the parameters of
the fits, except the radial peak positions, are sensitive to the total
amount of material in the IPT, whereas the radial positions of the
ribbon are determined by physical effects outside of the IPT
\cite<e.g.,>[see also \S\ref{future}]{barbosa83, ip83}.

We note that, as derived, the absolute values of the dawn ribbon
surface brightness values shown in Figure \ref{fig:time_sequence}
(bottom) are artificially low because, at the blueshift of the dawn
ribbon, IPT emission falls outside of the central bandpass of the
[S\;II]\,6731\,\AA\ filter as measured in a collimated beam.  This
effect can be corrected using a velocity-dependent IPT map, the
[S\;II] filter transmission curve, and consideration of the effects of
the telescope's F/11 light paths on the filter's total transmission.
However, making these corrections would not affect the results of our
current study.

\section{Results}
\label{results}

We anticipate the IoIO data will be very useful for correlative
  studies for observations focusing on Io and the effect that material
  escaping Io has on Jupiter's magnetosphere, such as afforded by
  NASA's \textsl{Juno} mission \cite{bolton17} and
  supporting observations \cite{orton_planets2020, orton22}.
To that effect, the surface brightness points shown in Figure
  \ref{fig:time_sequence} have been archived at Zenodo
  \cite{morgenthaler23_IoIO_data}.  In this paper, we
  focus on what the data themselves can say about the physical
  processes that drive material escape from Io's atmosphere.

Our 6-year time sequence of Jovian Na nebula and IPT
[S\;II]\,6731\,\AA\ ribbon brightnesses (Figure
\ref{fig:time_sequence}) shows considerable modulation in each
emission line as a function of time.  During each $\sim$7-month
observing window, at least 1 -- 2 enhancements, each lasting 1 -- 3
months are seen.  Very little time is spent in a quiescent state.
Visual inspection of Figure \ref{fig:time_sequence} reveals that the
average values of the Na nebula and IPT surface brightnesses are
determined by the enhancements, rather than any quiescent value.  To
quantify this finding we compute the \citeA{tukey77} biweight
distribution (Equation \ref{eq:biweight}) of the measurements
presented in Figure \ref{fig:time_sequence}.  Recall from
\S\ref{reduction:general} that the biweight distribution is a more
robust statistical measure of the central location of a distribution
than the median or average.  The biweight distribution values of the
Na nebula points are 80\,R, 50\,R and 15\,R in the 6\;R$_\mathrm{j}$,
12\;R$_\mathrm{j}$, and 24\;R$_\mathrm{j}$ apertures, respectively.
Compare this to low values of approximately 30\,R, 20\,R, and 5\,R.
Similarly, the biweight distributions of the ribbon brightnesses are
50\,R and 90\,R for dawn and dusk, respectively, with minima of 15\,R
and 30\,R.  

Visual inspection of Figure \ref{fig:time_sequence}
also shows a quasi-contemporaneous relationship between the Na
  nebula and IPT enhancements.  For instance, the relative timing
  between the peaks of the 2018 Na nebula and IPT enhancements is
  significantly different than that seen in 2020.  And the fall 2022
  Na nebula enhancement is particularly bright compared to other Na
  nebula enhancements, yet the IPT enhancement during that time period
  is particularly weak compared to other years.  This type of behavior
  has not been reported before.

We discuss the implications of our results in
  \S\ref{discussion}.  But first, we compare our results
  to those of previous studies, which provide valuable context to our
  discussions.

\section{Comparison to previous studies}
\label{compare}

IoIO occupies a unique niche in sensitivity, which ideally suits it to
study of the modulation of material flow from Io into the broader
Jovian magnetosphere.  The 35\,cm telescope aperture of IoIO was
chosen to be comparable to the smallest apertures that have
successfully imaged the IPT \cite{nozawa04_no_SII}.  This has allowed
us to reliably capture, at modest cost, a 6-year history of the
modulation in the IPT [S\;II] 6731\,\AA\ ribbon brightnesses
(presented here) and positions (to be presented in a subsequent work).
Our equipment choice limited the FOV of the instrument to
0.4$^{\circ}$, which is much smaller than the 2.5$^{\circ}$ --
7$^{\circ}$ FOV of long-term previous coronagraphic Na nebula studies
\cite{mendillo04, yoneda09, yoneda15, roth20}.  However, the narrower
FOV of IoIO affords it much greater sensitivity to emission close to
Io, as evident by comparing the left columns of our Figure
\ref{fig:Na_SII_images} to Figure 1 of \citeA{mendillo04} and Figures
2 of \citeA{yoneda09, yoneda15}.  This feature of the IoIO Na nebula
observations will allow us to conduct detailed morphological studies
of the jet and stream in future work.

\subsection{Sodium-related studies}
\label{Na_compare}

The outer portions of the IoIO FOV overlap with the inner portions of
the images recorded by other wide-field Na nebula studies
\cite{mendillo04, yoneda09, yoneda15, roth20}, allowing direct
comparison.  For instance, the peak intensity of the fall 2022 Na
nebula enhancement detected by IoIO roughly compares to the peak
intensities in the 2007 and 2015 enhancements captured by
\citeA{yoneda09, yoneda15}.  The \citeA{roth20} study is useful, since
it provides a time-history of Na nebula brightnesses measured with the
same coronagraph used in the \citeA{yoneda15} work for a 4-month
interval in 2017 during which no enhancement was reported.
Nevertheless, modulation at the $\sim$10\,R level in daily values
($\sim$5\,R in the half-month averages) is seen.  This is comparable
to the variation seen in the IoIO dataset during the 2018 enhancement.
The greater sensitivity of the IoIO coronagraph to emission closer to
Jupiter makes variation of this magnitude much easier to detect.  This
implies that periods formerly identified as quiescent in the
\citeA{yoneda15} dataset may, in fact, contain enhancements.  By that
interpretation, the period highlighted in the \citeA{roth20} appears
to be capturing the low point between two enhancements.

Spectroscopic observations conducted at the Lick observatory over the
entire 1995 Jovian opposition \cite{brown94T, brown97} are also useful
for comparison.  This work captured an enhancement in both the Na
5890\,\AA\ doublet  and [S\;II]
6731\,\AA\ (see also Sections
\ref{IPT_compare} -- \ref{contemporaneous_compare}).  The
10$^{\prime\prime}$ spectrograph slit was aligned along the
centrifugal equator, with peak emission averaged along the slit
reaching levels of 400\,R -- 800\,R.  In order to compare to our data,
we extend the aperture extraction procedure outlined in
\S\ref{reduction:Na} using apertures progressively closer to the
centrifugal plane following the same geometric sequence.  We stopped
decreasing the aperture size when the emission brightness increased by
$<$10\%.  The resulting aperture extended 0.5\;R$_\mathrm{j}$ above
and below the centrifugal plane and resulted in peak brightnesses of
200\,R -- 300\,R during the years 2017 -- 2021 and 630\,R in 2022.
This suggests that the 1995 Na enhancement captured by \citeA{brown97}
was comparable in size to the 2022 enhancement shown in Figure
\ref{fig:time_sequence} (upper panel).

Also important to mention are the \textsl{Galileo} dust detector
measurements acquired 1996 -- 2003 \cite{krueger03}.  There is
evidence that the dust comes from Io \cite{graps00,
  krueger03_graps00}, is composed almost entirely of NaCl
\cite{postberg06} and has its origin from Io volcanic plumes
\cite{krueger03}.  Further evidence shows that NaCl$^+$ is an
important pathway for Na escape from Io's atmosphere \cite{grava14,
  schmidt23}.  This suggests that variation seen in our Na nebula
dataset and others should be echoed in the \textsl{Galileo} dust
detector data.  \citeA{krueger03} used a simple geometric model of
dust emission from Io to translate dust detector count rates into the
flux of dust from Io (their Figure 2).  As discussed by
\citeA{krueger03}, the \textsl{Galileo} orbit precluded continuous
measurements of the dust streams before mid 2000.  However, beginning
after this time, there was a large, well-covered enhancement that
lasted $\sim$6 months.  Subsequent enhancements in the calculated Io
dust flux last $\sim$1 -- $\sim$3 months and have smaller amplitudes
than the 2000 enhancement.  The magnitudes of the enhancements are 1
-- 4 orders of magnitude, which is much larger than those seen in
sodium nebula data.  Full treatment of the reasons for the difference
in magnitude seen between the different measurement methods is beyond
the scope of this work.  Rather, we point out that the durations of
the enhancements in the derived dust flux from Io is comparable to
those observed in the Jovian sodium nebula \cite[and this
  work]{brown97, yoneda09, yoneda15}.

\subsection{IPT studies}
\label{IPT_compare}

Previous studies of IPT [S\;II] 6731\,\AA\ emission show peak ribbon
brightness values in the $\sim$100\,R -- $\sim$1000\,R range in
individual measurements \cite<e.g.,>{morgan85a, oliversen91,
  jockers92, woodward94, schneider95, thomas01, nozawa04_no_SII,
  yoneda09, schmidt18}.  As described in \S\ref{reduction:IPT}, the
values shown in Figure \ref{fig:time_sequence} are the average surface
brightness of each ribbon derived using a two-step Gaussian fitting
procedure, with one Gaussian used to isolate emission in the radial
direction and one to compute the average surface brightness in the
vertical.  To convert from averages over the Gaussian functions to
peak values, we multiply by $2 \pi$, one factor of $\sqrt{2\pi}$ for
the integral over the vertical Gaussian and another factor of
$\sim\sqrt{2\pi}$ to account for the summation between the $\pm 1
\sigma$ limits of the radial Gaussian.  Following our 21-point moving
averages, this yields peak values of $\sim$200\,R -- $\sim$900\,R,
with individual points ranging from $\sim$50\,R to $\sim$1200\,R.  We
take this to be good agreement with previous studies and therefore
independent validation of our stellar calibration procedure.

The only other published study of IPT [S\;II] 6731\,\AA\ emission
lasting more than a few weeks during a single Jovian opposition is the
companion of the spectroscopic Na nebula observations collected in
1995 at the Lick observatory, discussed in \S\ref{Na_compare}
\cite{brown94T, brown97}.  That study captured an IPT enhancement that
lasted $\sim$2.5 months.  The emission pre- and post enhancement was
$\sim$200\,R and the emission was $\sim$400\,R at its peak.  The
factor of $\sim$2 difference between the pre/post and peak values is
comparable to the broad enhancement partially captured at the
beginning of the 2019 Jovian opposition.  In other words, within the
range observed over our 6-year study.

Also useful for comparison are the two long-term studies of the warm
torus that have been conducted in the EUV, one by \textsl{Cassini} and
one by \textsl{Hisaki} \cite<e.g.,>{steffl04a, yoshikawa17}.
Comparison of the surface brightnesses seen in the EUV warm torus
observations to the surface brightness of the [S\;II]
6731\,\AA\ observations of the ribbon region would require detailed
IPT modeling that is beyond the scope of this work.  Thus, we limit
our discussion to the duration of the enhancements.  During the
duration of its observations, \textsl{Cassini} captured two
enhancements in emissions from ionization states of S and O lasting of
order 1 -- 3 months, one in late 2000, the other in early 2001.
During its multi-year observing campaign \textsl{Hisaki} saw one large
enhancement that lasted $\sim$3 months in 2015.  Smaller amplitude
modulations in the \textsl{Hisaki} data have also been noted
\cite{roth20}.  These enhancement durations are comparable to those
seen in the IoIO data.

\subsection{Contemporaneous Na nebula and IPT studies}
\label{contemporaneous_compare}

Two previous studies reported contemporaneous Na nebula and IPT
  enhancements \cite{brown97, tsuchiya18}.  These
  studies, and related work, concentrated on the detailed behavior of
  the observed emission during the enhancements and the implications
  for physical processes occurring within the IPT and broader Jovian
  magnetosphere \cite<e.g.,>{yoshikawa17, kimura18, hikida18,
    yoshioka18, tsuchiya18, hikida20, roth20, tao16a, tao16b, tao18,
    tao21}.  Such in-depth study of individual enhancements is
  beyond the scope of our current work.  Rather, we note for
  comparison to our data, that for both the 1995 and 2015
  enhancements, there was delay of $\sim$4 weeks between the peak in
  the Na nebula and IPT S$^+$ emissions, even though in the 2015 case,
  the S$^+$ emissions were from the ribbon region and detected via the
  [S\;II] 6731\,\AA\ line and in the 2015 case, the
  emissions were from the warm torus and detected via the  S\;II
  765\,\AA\ line.  Because different regions of the torus were
    studied in the two cases, comparison of the relative strengths of
    the IPT enhancements requires modeling that is beyond the scope of
    this effort.  Thus, we are not currently able to use these studies
    to corroborate our observation that the relative strengths of the
    Na nebula and S$^+$ enhancements can vary significantly with
    time.

\section{Discussion}
\label{discussion}

Our study has a unique combination of sensitivity, cadence and
duration that has enabled it to determine that the Na nebula and IPT
are frequently in states of enhancement and that the enhancements
  in the two species have a quasi-contemporaneous relationship.  When
  interpreted within the context of previous studies, the former
  result allows us to rule out solar insolation-driven sublimation as
  the primary mechanism driving Io atmospheric escape; the later
  provides insights into the most likely mechanism -- volcanism -- and
  the subsequent path sodium- and sulfur-containing materials take
  through and out of Io's atmosphere.  Our discussion begins with
  atmospheric escape.  

\subsection{Response of sodium nebula and IPT to Io
    atmospheric escape}
\label{escape}

As reviewed in \S\ref{introduction}, Io's
  atmosphere is removed by interaction with the IPT via charge
  exchange, sputtering, and electron impact ionization to fill the
  neutral clouds on timescales of hours \cite<e.g.,>{smyth88,
    dols08, dols12, smith22}.  The apertures used to extract Na
  nebula surface brightnesses (Figure \ref{fig:Na_apertures})
are primarily filled by sodium traveling near IPT's
  $\sim$70\,km\,s$^{-1}$ corotation velocity.  The residence time of
  this material in the IoIO FOV is $\sim$11 hours.  Furthermore, we
  have chosen the apertures that integrate over the effect of
  Jupiter's $\sim$10\,hr rotation period and Io's $\sim$40\,hr orbit.
  Thus, to the accuracy of $\sim$1 day, the Na nebula surface
  brightnesses shown in Figure \ref{fig:time_sequence}
(top panel) provide a good indicator of the modulations in the
  escape rate of sodium-bearing material from Io's atmosphere.

The response of the IPT to Io atmospheric escape is somewhat more
  complicated than that of the Na nebula.  The neutral clouds
  described in the previous paragraph are shaped by interaction with
  the IPT through the processes of impact ionization and
  charge-exchange \cite<e.g.,>{smyth03}. Impact
  ionization results in the addition of new material and proceeds on
  timescales of $\sim$ 1 day \cite{smyth03}. The
  residence time of plasma in the IPT, is 20 -- 80 days, with the
  shorter residence times corresponding to times of higher total
  plasma density \cite{bagenal11, hess11, copper16,
    tsuchiya18}.  Thus, when there is an enhancement in the
  escape of sulfur-bearing material from Io's atmosphere, the peak in
  the IPT S$^+$ 6731\,\AA\ ribbon will lag by an amount
  dependent on the IPT plasma density.  A model is being developed
  that could, in principle, calculate the precise IPT ribbon response
  \cite{coffin20, coffin22AGU, coffin23AGU, nerney20,
    nerney22AGU, nerney23DPS}, but its completion and
  application to the IoIO dataset is beyond the scope of our current
  project.  Thus, we take the $\sim$4 week delay between the peaks in
  Na nebula and IPT S$^+$ emission in the previous two studies that
  captured contemporaneous enhancements
  \cite<\S\ref{contemporaneous_compare};>{brown97,
    tsuchiya18} as indicative of the plasma transport time
  during a typical IPT enhancement.  Four weeks is also similar to the
  transport time deduced from the larger IPT enhancement captured by
  \textsl{Cassini} \cite{steffl04a, copper16}.


\subsection{Interpretation of quasi-contemporaneity of Na
    nebula and IPT enhancements}
\label{quasi}

Within the context of the discussion in Sections
  \ref{contemporaneous_compare} and \ref{escape}, we can
  now offer an interpretation the quasi-contemporaneous nature of the
  Na nebula and IPT enhancements seen in Figure
  \ref{fig:time_sequence}. In all cases except 2020,
  each major Na nebula enhancement has a companion enhancement seen in
  the IPT nebula that is delayed by $\sim$4 weeks.  This $\sim$4 week
  delay is consistent with that seen in previous studies and is
  indicative of simultaneous release of sodium- and sulfur-bearing
  material from Io's atmosphere.  In 2020, the delay between the Na
  nebula and IPT enhancements peaks is almost twice as long, however,
  the profiles of both enhancements are more complicated than the
  enhancements in other years: the Na nebula enhancement has a
  shoulder on its trailing edge and the IPT enhancement appears to
  consist of a broad, main peak, followed by a small, sharp peak.
  Thus, we offer the suggestion that in mid 2020, there are two
  overlapping sets of Na nebula and IPT enhancements, with the earlier
  set being larger than the later.  A similar relationship may exist
  between other, smaller enhancements, such as the shoulder on the
  early 2020 Na nebula peak and the small IPT peak in mid 2019.  We
  discuss the implications of the variation in the relative Na nebula
  and IPT peak \emph{sizes} seen in each contemporaneous pair (e.g.,
  fall 2022 being the most extreme) in
  \S\ref{ratio_modulation}.

\subsection{Ruling out solar insolation-driven
  sublimation}
\label{sublimation}

The current paradigm holds that the bulk of the escaping material
  from Io's atmosphere is supplied by Io's global sublimation
  atmosphere \cite<e.g.,>{schneider07, dols08, dols12}.
This paradigm suggests that, in the absence of some
  other perturbing effect on Io's atmosphere, the variations seen in
  the Na nebula and IPT should be dominated by variations in solar
insolation, that is, Io's 42\,hour orbit or Jupiter's 12\,year orbit
\cite<e.g.,>{depater20_post_eclipse, tsang12}.  Enhancement in the
escape of material from Io's atmosphere may also be modulated by
Jupiter's magnetic rotational period (9.925\,hr) due to Io's apparent
motion within the IPT \cite<e.g.,>{smyth11}.  Even when
  considering the timescales of the responses of the Na nebula and IPT
  to material escape from Io's atmosphere, discussed in 
  Sections \ref{escape} -- \ref{quasi}, solar insolation and
  magnetic periodicities are not compatible with the behavior of the
  enhancements seen in Figure \ref{fig:time_sequence}.
  Thus, we
  argue that one or more other physical mechanisms are
driving atmospheric escape during
  enhancements.  Since enhancements dominate the average supply of
  material from Io's atmosphere (Sections \ref{results},
  \ref{escape}), the process(es) driving enhancements provide the
  bulk of the material in the Jovian sodium nebula and IPT.

\subsection{The case for volcanism}
  \label{volcanism}

The initial claim of a link between Io volcanism and material
  release from Io's atmosphere was made using Jovian sodium nebula
  images recorded with a cadence of weeks to months and
  disk-integrated infrared observations \cite{mendillo04}.
A subsequent study using Jovian sodium nebula observations
  recorded with a near-nightly cadence and a much more extensive set
  of disk-resolved Io infrared observations, has failed to validate
  that initial claim \cite{roth20}.  In this work, we
  take a different approach and use the time behavior of material
  release from Io's atmosphere to suggest the most likely driver for
  atmospheric escape is volcanic plumes.

Recall from \S\ref{introduction}, that based on EUV brightness of the
IPT, $\sim$1\,ton\,s$^{-1}$ of material must be flowing into it from
Io's atmosphere.  We can now attribute this amount of flux to the
mechanism(s) that cause enhancements.  Io's atmosphere is itself
tenuous and, without resupply from the surface or subsurface,
cannot act as a reservoir for supplying enhancements in escape that
last weeks to months.  We have ruled out variation due to solar
insolation-induced sublimation and magnetospherically enhanced escape
of Io's global atmosphere, in \S\ref{sublimation}.  Thus,
geologic processes of some sort must ultimately be involved in the dominant
  process(es) of atmospheric escape from Io's atmosphere.

In \S\ref{quasi}, we showed that enhancements in
  the escape of sodium and sulfur-bearing material from Io's
  atmosphere occur simultaneously.  This simultaneity, together with
  the geologic nature of the processes driving escape imply that there
  is a single geographical location responsible for driving
  atmospheric escape in each pair of enhancements, with that location
  not necessarily being the same for each pair.  Finally, we note that
  behavior of the Na nebula and IPT surface brightnesses appears to be
  stochastic, with large and small enhancements interleaved.  The
  picture that emerges is that geologic activity at discrete sites on
  Io results in enhancements in the escape of sodium- and
  sulfur-bearing materials that last 1 -- 3 months, with 1 -- 2
  enhancements seen during each 7 month Jovian opposition observing
  window and with the relative amount of material escape varying
  between individual events.

Of the known geologic processes on Io that match the above
  criteria, volcanism is the most likely to result in the
stochastic perturbation of
Io's atmosphere, via processes involving plumes.   Observational support for Io
  volcanic plume-driven atmospheric escape comes from the correlation
between plumes observed by \textsl{Galileo} \cite{geissler04_surf,
  geissler08} and enhancements in the Jovian dust streams
\cite{krueger03}.  As reviewed in \S\ref{Na_compare}, the dust streams
are composed almost entirely of NaCl and come from Io
\cite{graps00,krueger03_graps00,postberg06}.  Furthermore, NaCl$^+$
has been shown to be an important pathway for Na escape from Io
\cite{grava14,schmidt23}.  Based on this evidence, it is plausible to
suggest that volcanic plumes play a key role in the supply of the
Jovian sodium nebula.  We use the very large Jovian dust stream
  enhancement that peaked in early September 2000 and the IPT
  enhancement observed by \textsl{Cassini} that peaked about a month
  later as observational evidence of the connection between IPT
  enhancements and volcanic plumes \cite[see also
    \S\ref{IPT_compare}]{krueger03, steffl04a}.

The difficulty with suggesting that volcanic plumes themselves are
responsible for launching  material out of Io's atmosphere
is that plume vent velocities are far below Io's gravitational escape
velocity, even when considering local atmospheric heating by plume
dynamics \cite<e.g.,>{schneider07, mcdoniel17, depater20_post_eclipse,
  redwing22}.  Thus, if plumes are implicated in material escape from
Io, the mechanism must be indirect.  Plume models show that shocks in
the plume canopy impede upward flow of material, redirecting it
outward and toward the surface \cite{zhang03, zhang04}.
\citeA{zhang03} suggest that the material that is redirected forcibly
toward the surface enhances sublimation of SO$_2$ frost over a large
area.  Sublimation of SO$_2$ frosts by hot surface/subsurface lavas
would provide a similar localized enhancement in Io's SO$_2$
atmosphere.  This SO$_2$ would then be available to interact with the
IPT via the pathways of sputtering, electron impact ionization and
change exchange.  The difficulty with this scenario is that it does
not provide a comparable mechanism for enhancing the escape of NaCl,
since NaCl has a much higher sublimation temperature than SO$_2$.
However, the amount of NaCl provided by the plume itself and/or NaCl
lofted from Io's surface by sublimation may be sufficient to drive
escape, when considering interaction with the IPT over a large area
(see also \S\ref{atmosphere}).

Perhaps more a more plausible path of escape for both SO$_2$ and NaX
is to consider the ability of plumes to loft material to sufficient
altitude to enable enhanced interaction between IPT plasma and the
tops of plumes.  An extension of the \citeA{zhang03, zhang04} model of
Pele's plume shows that by including interaction between the top of
the plume and the IPT, better agreement is found with the distribution
of material seen on the surface \cite{mcdoniel15, mcdoniel17,
  mcdoniel19}.  The \citeA{mcdoniel19} work provides theoretical
validation of the ability of plume tops and IPT plasma to interact,
but stops short of a full quantitative calculation of the amount of
material that could be removed by this interaction.  Thus, although there is observational
  evidence that connects volcanic plumes to enhancements in the escape
  of sodium- and sulfur-bearing material from Io's atmosphere, current
  state-of-the-art theoretical calculations have not been able to
  determine the exact pathway taken by the material to Io's
  exosphere.

\subsection{Implications for the transport of sodium- and
    sulfur-bearing material through Io's atmosphere}
\label{atmosphere}

Atacama Large Millimeter/submillimeter Array (ALMA) observations
  of Io's atmosphere reveal collections of hot NaCl, KCl and SO$_2$
  gases, that are interpreted as plumes \cite{redwing22}.
  \citeauthor{redwing22} note that the highest column density
  collections of alkali and SO$_2$ gasses are consistently \emph{not}
  found to be coincident with each other (Figure 9 of that work).  As
  discussed by \citeauthor{redwing22}, these results are
  difficult to explain given that SO$_2$, the primary volatile on Io,
  is expected to be associated with all plume activity.  These results
  are also in apparent conflict with our result that sodium- and
  sulfur-bearing material are consistently seen to escape at the same
  time, implying a common geographic source (Sections
  \ref{quasi} -- \ref{volcanism}).  In this Section, we
  discuss some mechanisms that might contribute to this effect,
  including those not considered by \citeauthor{redwing22}.

\citeA{redwing22} provide two potential
  reasons for the lack of spatial coincidence between alkali and
  SO$_2$ plumes in Io's atmosphere which rely on the difference in
  vaporization pressures of these materials.  (1) SO$_2$ gas is
  produced primarily by hot lava vaporizing frost deposits, with these
  deposits found primarily at low- to mid-latitudes.  Alkalis sublime
  at a much higher temperature and therefore will not be released into
  Io's atmosphere by this effect.  (2) The alkalis observed by ALMA
  are released in the plumes of high-temperature volcanoes.
  \citeauthor{redwing22} note that these
  high-temperature plumes should also produce SO$_2$ but that these
  alkali-producing volcanoes are consistently located at high
  latitudes where atmospheric temperatures may be low enough to freeze
  SO$_2$ within the plumes.  In this way, (1) explains why SO$_2$
  plumes appear in the absence of alkali plumes (low latitudes) and
  (2) suggests that SO$_2$ is always collocated with the high-latitude
  alkali plumes, however, this high-latitude SO$_2$ is largely
  invisible to ALMA because it is in solid form.  Evidence of
  solid-phase transport of SO$_2$ through Io's atmosphere comes from
  \textsl{Galileo} detection of very high mass-to-charge ratio ions,
  interpreted as clusters of SO$_2$ molecules, or ``snowflakes,'' when
  it flew over Io's north pole \cite{frank02}.  This
  explanation requires that the SO$_2$ gas in the plumes at low
  latitudes not contribute significantly to SO$_2$ escape, and
  requires NaCl to be primarily sourced from the polar plumes,
  possibly by the mechanism discussed in the next paragraph.  Plume
  models have yet to be constructed that would test this
  ``snowflakes'' hypothesis.

Another way to ``hide'' SO$_2$ and/or NaCl from ALMA is to ionize
  them.  The behavior of Io's auroral Na, O, SO$_2$, and SO emission
  while Io transitions to and from eclipse behind Jupiter provides
  evidence that (a) photoionization is the primary mechanism for
  producing SO$_2^+$ and (b) SO$_2^+$ plays an important role on the
  pathway of NaCl escape from Io's atmosphere via charge-exchange
  \cite{schmidt23}.  Io's polar atmosphere is exposed to
  the sun for longer periods of time than that above the rest of Io,
  thus increasing the average rate of photoionization at the poles.
  Furthermore, Io's collisional atmosphere is thinner in these regions
  \cite<e.g.,>{walker10}, providing more access of plume
  material to the exosphere, where IPT-driven escape processes are the
  most efficient.  This suggests that plumes at the poles will have
  enhanced escape (see also \S\ref{ratio_modulation}).
Like the SO$_2$ ``snowflake'' atmospheric transport mechanism
  suggested in the previous paragraphs, this mechanism favors the
  NaCl-rich high-latitude plumes detected by ALMA as the source of the
  sodium- and sulfur-bearing material contributing to the Na nebula
  and IPT enhancements.

For logical completeness, we also consider the suggestion that
  NaCl-containing volcanic dust and ash (``dust bunnies'') may
  transport NaCl through Io's atmosphere in a form not visible to
  ALMA.  To simultaneously explain the ALMA and IoIO data, this would
  imply that these large particles are driven through the atmosphere
  by the SO$_2$ plumes at low latitudes and that those particles are
  quickly charged by interaction with the IPT and removed from ALMA's
  FOV by Jupiter's magnetic field.  Importantly, under this
  interpretation, the high-latitude NaCl plumes seen by ALMA would not
  contribute to escape and the SO$_2$ from these plumes must still be
  invisible to ALMA, necessitating one or more of the mechanisms in
  the previous paragraphs.


In support of the ``dust bunny'' hypothesis, we recall the
  observational evidence we used to connect plumes to the Jovian
  sodium nebula \cite<\S\ref{volcanism};>{krueger03, grava14,
    schmidt23}.  \citeauthor{krueger03} found that dust stream
  enhancements were likely associated with the plumes of volcanoes
  such as Pele, Tvashtar, a region near the north pole, and a region
  south of Karei \cite<now known as Grian;>{geissler08}.
The plume deposits of these eruptions are primarily SO$_2$ rich,
  with minor contributions from silicate ash \cite{geissler99,
    geissler04_surf}.  The association between atmospheric
  escape of sodium- and sulfur-bearing material and these large,
  SO$_2$-rich plumes is suggestive that the large atmospheric
  disturbances caused by these plumes may be the key to driving escape.
  The lack of association between SO$_2$-dominated plumes and NaCl
  emission in the ALMA data then becomes support for transport of NaCl
  through the atmosphere in these plumes in a form such as dust or
  ash.

Finally, we suggest that there may be no need to ``hide'' SO$_2$
  from ALMA in the regions where bright NaCl emission is seen.  In the
  ALMA observation of NaCl and SO$_2$ that has the highest resolution,
  relatively faint concentrations of SO$_2$ gas are seen in the
  vicinity of the brightest NaCl emission \cite<see 2016-07-26
  in Figure 9 of>{redwing22}.  More detailed analysis of the
  ALMA data would be needed to determine if the NaCl and SO$_2$ seen
  in this observation is consistent with a single geographic source.
  Plume models could be used to determine if the amount of SO$_2$ is,
  in fact, less than what is expected for the range volcanic activity
  that has been observed on Io \cite<see, e.g. review
  by>{depater21}.  Additional high-resolution ALMA images,
  ideally recorded contemporaneously with IoIO data, would also be
  useful.  If the NaCl to SO$_2$ ratio were to be found to be
  reasonable in NaCl plumes, this could imply that atmospheric escape
  is primarily tied to NaCl-rich and/or high-latitude plumes and that
  the collections of high column density SO$_2$ gas seen at lower
  latitudes in the ALMA data contribute at most a minor, steady amount
  to the IPT (e.g., baseline in Figure
  \ref{fig:time_sequence}, bottom panel.)


\subsection{Processes that modulate Na nebula and IPT
    enhancement sizes}
\label{ratio_modulation}

Having established that volcanic plumes are the likely
  precipitating agent for material escape from Io's atmosphere and
  discussed possible pathways of material transport through Io's
  atmosphere, we return to our discussion of the quasi-contemporaneous
  nature of the Na nebula and IPT enhancements, begun in
  \S\ref{quasi}, to offer possible causes for the
  modulation seen in the \emph{sizes} of Na nebula and IPT
  enhancements.  These suggested causes divide into four general
  categories: (1) variation in the content of sodium- and
  sulfur-bearing material in volcanic plumes, such as produced by
  different magmas \cite<e.g., see>{redwing22}; (2)
  processes involving the interaction of the plumes with the
  atmosphere, such as shocks
  \cite<\S\ref{volcanism};>{zhang03, zhang04} (3)
  modulation in material transport through the atmosphere
  \cite<\S\ref{atmosphere};>{schmidt23}; and (4)
  variation in the efficiency of escape.  In this Section, we
  concentrate on category (4), because understanding effects in this
  category greatly enhances the ability to use Jovian sodium nebula
  and IPT data to make progress understanding physical effects in
  categories (1) -- (3).

One of the scenarios discussed in \S\ref{atmosphere}
suggested that enhanced SO$_2$ ionization in Io's polar regions
  may play a role in enhancing NaCl escape \cite{schmidt23}.
This would favor the Tvashtar plume (63$^\circ$N 124$^\circ$W)
  over that of Pele (19$^\circ$S 255$^\circ$W) or Surt (45$^\circ$N,
  336$^\circ$W) for the source of the very large Jovian dust stream
  enhancement observed by \textsl{Galileo} in late 2000, even though
  Surt, active at the time, had a much larger infrared output and Pele
  was the most active large plume during the \textsl{Galileo} era
  \cite{marchis02, porco03, krueger03, geissler04_surf}.
Because SO$_2$ ionization and subsequent rapid dissociation
  \cite{huddleston98} may be enhanced in the polar
  regions \cite{schmidt23}, the efficiency of S and O
  production, and thus escape, may be enhanced at the poles as well.

Because NaCl$^+$ has an important role in the pathway of Na
  escape from Io's atmosphere \cite<Sections \ref{Na_compare},
  \ref{volcanism}, \ref{atmosphere};>{schmidt23}, sodium-bearing
  material in Io's anti-Jovian equatorial exosphere may have an
  exaggerated escape efficiency over that of sulfur-bearing material.
  The sodium ``jet'' was seen to be rooted in the anti-Jovian
  equatorial region of Io's exosphere the one time it has been imaged
  in sufficient spatial detail for detection near Io
  \cite{burger99}.  \citeauthor{burger99} offered two
  hypotheses to explain this behavior: (i) Material was being injected
  into the exosphere from below at this location, e.g., by volcanism.
  (ii) Ionization is enhanced in this region due to Io equatorial
  auroral activity \cite<e.g.,>{roesler99, geissler99,
    roth14}.  Case (ii) would provide a mechanism for enhancing
  material flow into the jet, over material flow to the IPT, since the
  jet is formed by prompt neutralization of pickup ions within Io's
  exosphere \cite{schneider91Corona, wilson94}, whereas
  direct ionization of material in Io's exosphere has been shown to be
  a minor contributor to the influx of plasma to the IPT
  \cite<\S\ref{introduction};>{dols08,
    dols12}.  The prompt neutralization of the sodium forming
  the jet also explains why it is not expected to be seen rooted at
  the sub-Jovian auroral spot: the initial gyration of the ions
  directs them into Io's surface.  We note that hypotheses (i) and
  (ii) are not necessarily mutually exclusive: the ``jet'' may always
  be rooted in the region of the anti-Jovian equatorial auroral spot,
  but its response to atmospheric escape may be exaggerated if that
  atmospheric escape is located in that region.

Finally, dust, which acquires a negative charge
  \cite{zook96}, enhances the initial escape of
  Na-bearing material on Io's sub-Jovian hemisphere
  \cite{grava21}.  As the dust particles are destroyed
  and the constituent molecules and atoms are released, they acquire a
  positive charge and join the ``jet'' feature.  A positive
  identification between a Na nebula enhancement and a plume on the
  sub-Jovian hemisphere, could thus potentially be used in support of
  the ``dust bunnies'' hypothesis discussed in
  \S\ref{atmosphere}.

\subsection{Toward a connection with Io infrared
    observations}
\label{IR}

Since the early 1980s, synoptic Io infrared observations have
  been used to help understand Io volcanic processes and their
  geologic implications \cite<e.g., see review
  by>{depater21}.  As noted in \S\ref{volcanism},
these infrared observations have been compared to sodium nebula
  observations in an attempt to establish a correlation
  \cite{mendillo04} which has not stood the test of time
  \cite{roth20}.  The initial attempt to connect
  infrared indicators of Io volcanic activity to enhancements in the
  sodium nebula made the implicit assumption that the brightest
  infrared events should be correlated to the brightest nebula
  images.  Here we suggest that instead, the dimmer infrared events
  may be more likely to be correlated to the brighter sodium nebula
  enhancements. 

One of the fundamental results of our study is that the Jovian
  sodium nebula and IPT show contemporaneous enhancements of varying
  relative amplitudes that last 1 -- 3 months (Figure
  \ref{fig:time_sequence}; Sections \ref{results},
  \ref{escape} -- \ref{quasi}).  A long-term study of the time
  variability of Io's hotspots found that they divided into two
  groups: those with persistent activity and those that exhibited
  sudden brightening, followed by a steady decay
  \cite{dekleer16}.  For those hotspots that exhibited
  sudden brightening events, the brighter the event, the shorter the
  decay \cite<see, e.g., Figure 11 of>{dekleer16}.  The
  hot spots with decay times of order 1 month or longer were dimmer,
  by a factor of 5 or more, than the brightest outbursts.  If we make
  the very simplistic assumption that in a volcano where plume
  activity is found, plume activity will persist over roughly the same
  time period as infrared activity, we can support the argument that
  infrared hotspots exhibiting eruption phases lasting 1 -- 3 months
  are the more likely to be correlated to atmospheric release events.
  Thus, dim infrared outbursts may be the most likely to be
  correlated with enhancements in material release from Io's
  atmosphere.

\section{Summary and Conclusions}
\label{conclusion}

We have used IoIO, an observatory composed almost entirely
off-the-shelf equipment \cite<\S\ref{observations}
and>{morgenthaler19} to collect the largest set of contemporaneously
recorded Jovian sodium nebula and Io plasma torus (IPT) in
[S\;II]\,6731\,\AA\ images assembled to date (see examples in Figure
\ref{fig:Na_SII_images} and accompanying animations).  Using simple
image analysis techniques (\S\ref{reduction}), we construct a time
history of the brightnesses of the Na nebula and IPT [S\;II] emission
(Figure \ref{fig:time_sequence}).  Qualitative inspection of this
Figure shows 1 -- 2 enhancements in the Na nebula and IPT [S\;II]
emission per $\sim$7-month observing window, such that a quiescent
state of emission is rare (\S\ref{results}).  The minimum and
  maximum surface brightness values seen in the IoIO Na nebula and IPT
  images compare favorably with previous studies
  (\S\ref{Na_compare} -- \ref{IPT_compare}).  Most large
  IPT enhancements peak $\sim$4-weeks after the corresponding
  enhancement in the Na nebula, as seen in previous studies
  (\S\ref{contemporaneous_compare}) and as expected from
  plasma transport within the IPT (\S\ref{escape}).  The
  exception to this, seen in mid 2020, is likely caused by the overlap
  of multiple enhancements (\S\ref{quasi}).  We rule out
  sublimation as the primary driver of material escape from Io's
  atmosphere in \S\ref{sublimation}.  This is our
  most definitive result.

Having ruled out sublimation as the primary driver of atmospheric
  escape from Io, we show that geologic activity in some form, likely
  volcanic plumes, drives escape, either directly or indirectly
  (\S\ref{volcanism}).  In light of other published
  results, this has implications on the transport of material through
  Io's atmosphere (\S\ref{atmosphere}).  In
  \S\ref{ratio_modulation}, we review the processes that
  can modulate the relative sizes of contemporaneous Na nebula and IPT
  enhancements, focusing on processes that might modulate the
  efficiency of Io atmospheric escape as a function of geographic
  location.  Finally, in \S\ref{IR}, we note that Io's
  dimmer infrared outbursts have durations and time profiles similar
  to Na nebula and IPT enhancements, suggesting that it these 1 -- 3
  month-long infrared outbursts may be the more likely to show
  correlation with the release of material from Io's atmosphere.

In conclusion, our work shows that off-the-shelf equipment with
  minimal customization, together with simple analysis techniques can
  be used to collect data that provides valuable insights into the
  processes which produce material on Io's surface, transport it
  through its atmosphere, and release it into Jupiter's broader
  magnetospheric environment.

\section{Future Plans}
\label{future}

We have pointed out the existing observational evidence that
  links plume activity on Io to atmospheric escape in
  \S\ref{volcanism}.  Further confirmation of this link
  may be accomplished by accumulating additional contemporaneous IoIO
  observations of the Na nebula and IPT together with ALMA
  observations of Io's atmosphere -- currently there is only overlap
  during March 2018 \cite{redwing22}.  Disk-integrated
  observations conducted by the NOrthern Extended Millimetre Array
  (NOEMA) interferometer of the Institut de Radioastronomie
  Millimetrique (IRAM), such as those conducted by
  \citeA{roth20}, may also be useful.  Continued
  theoretical work on the effect that plumes have on Io's atmosphere
  and exosphere, as well as the interaction between the exosphere and
  the IPT is also needed \cite<e.g.,>{bloecker18, mcdoniel19,
    dols08, dols12, dols23, adeloye23DPS}.  These observational
  and theoretical studies can also be useful to help differentiate
  between the hypotheses offered in \ref{atmosphere}
concerning material transport through Io's atmosphere.

Continued disk-resolved observations of Io IR activity, such as
those carried out at the Keck and Gemini telescopes
\cite{dekleer19b} will also be interesting as
they might lead to validation of the correlation that was initially
claimed between Na nebula, IPT and IR brightnesses \cite{mendillo04,
  yoshikawa17, yoshioka18, tao18, koga18_time}, but has subsequently proven
elusive \cite{dekleer16AGU, roth20}.

We are also planning to conduct more detailed analysis of the
  IoIO images.  For instance, the IoIO images of the Na nebula contain
  three distinct features -- the ``banana,'' ``jet,'' and ``stream''
  -- that can be used to estimate the neutral sodium source rate from
  Io \cite{wilson02}.  The IPT ribbon positions, which
are detectable with IoIO (Figure \ref{fig:SII_reduction}, lower
panels), are related to the dawn-dusk electric field, which is
modulated by a combination of material flow toward the magnetotail and
solar wind pressure \cite{barbosa83, ip83}.  When combined with the
analysis presented here, the IPT ribbon positions retrieved from the
IoIO data will provide a significant amount of information regarding
the production of material on Io and its subsequent flow through and
out of Jupiter's magnetosphere.  


Finally, the unique sensitivity of IoIO to Na nebula and IPT
[S\;II]\,6731\,\AA\ enhancements, together with reliable robotic
operation and $<$24 hour turnaround for pipeline reduction ideally
suits it to provide real-time alerts of enhancements in the departure
of material from Io's atmosphere.  These can inform planned
observations of Io from both ground- and space-based platforms.  In
particular, nearly all of the plasma found in Jupiter's magnetosphere
comes from Io and makes its way through the IPT in 20 -- 80 days
before rapidly spiraling out through the rest of the magnetosphere
\cite{bagenal11, hess11, copper16, tsuchiya18}.  The modulations seen
in the lower panel of Figure \ref{fig:time_sequence} therefore precede
modulation in plasma density throughout the Jovian magnetosphere, a
feature that can be used to enhance the science operations of the
\textsl{Juno} mission \cite{bolton17} and supporting observations
\cite{orton_planets2020, orton22}.  NASA's \textsl{Europa Clipper}
\cite{howell20} and ESA's \textsl{JUICE} \cite{grasset13} missions
will  benefit from planned IoIO observations,
because of the record of exogenic material impinging on Europa,
Ganymede and Callisto during those missions.  Also, because
enhancements in the Jovian dust streams can induce detector fatigue in
\textsl{Europa Clipper's} SUrface Dust Analyzer \cite<SUDA;>{goode23},
IoIO observations can be used to inform SUDA operations while
\textsl{Europa Clipper} is sampling the broader Jovian magnetospheric
environment and thus optimize detector performance for that mission's
primary target.

\section*{Open Research Section}

The following software was used in this project: Astrometry.net
\cite{lang10}, AstroPy \cite{astropy_collaboration22}, Astroquery
\cite{ginsburg19}, BigMultiPipe \cite{morgenthaler22bmp}, Burnashev
\cite{morgenthaler23burnashev}, CCDMultiPipe
\cite{morgenthaler23ccdmultipipe}, ccdproc \cite{ccdproc17}, IoIO
control software \cite{morgenthaler23IoIO_code}, matplotlib
\cite{matplotlib}, moviepy \cite{moviepy21}, NumPy
\cite{oliphant06, numpy20}, photutils \cite{photutils22},
precisionguide \cite{morgenthaler23precisionguide}, Python 3
\cite{python3}, reproject \cite{robitaille20}, SciPy
\cite{SciPy20}, specutils \cite{specutils22}

The reduction products used to create Figure \ref{fig:time_sequence}
are archived with Zenodo \cite{morgenthaler23_IoIO_data}.


\acknowledgments

IoIO is hosted at the San Pedro Valley Observatory, near Benson,
Arizona, and has benefited greatly from the services provided by
observatory manager, Dean Salman.  Gavin Nelson, Andy Sorenson,
Elisabeth Adams, and the staff of Starizona have also provided
invaluable assistance in observatory maintenance.  The authors
  are grateful for the insightful comments received during the
  manuscript review process. This work has been supported by NSF
grants 1616928 and 2109219 and NASA grant 80NSSC22K0317 to the
Planetary Science Institute and NASA grant 80NSSC20K0559 to Boston
University.

%
%

\bibliography{jpmorgen_bibliography.bib}

%
%
%
%
%

\end{document}